\documentstyle[12pt]{article}
\begin{document}
\def\thefootnote{\fnsymbol{footnote}}
\begin{flushright}
KANAZAWA-06-12  \\
September, 2006
\end{flushright}
\vspace*{2cm}
\begin{center}
{\LARGE\bf Neutrino masses and CDM in a non-supersymmetric model}\\
\vspace{1 cm}
{\Large Jisuke Kubo}
\footnote[1]{e-mail:~jik@hep.s.kanazawa-u.ac.jp}
~{\Large and }~
{\Large Daijiro Suematsu}
\footnote[2]{e-mail:~suematsu@hep.s.kanazawa-u.ac.jp}
\vspace {1cm}\\
{\it Institute for Theoretical Physics, Kanazawa University,\\
        Kanazawa 920-1192, Japan}\\
\end{center}
\vspace{1cm}
{\Large\bf Abstract}\\
We propose a model for neutrino mass generation based on 
both the tree-level seesaw mechanism with a single right-handed neutrino 
and one-loop radiative effects in a non-supersymmetric framework.
The generated mass matrix is composed of two parts which have 
the same texture and produce neutrino mass eigenvalues 
and mixing suitable for the explanation of neutrino oscillations. 
The model has a good CDM candidate which contributes to the radiative 
neutrino mass generation. 
The stability of the CDM candidate is ensured by
$Z_2$ which is the residual symmetry
of a spontaneously broken U(1)$^\prime$.
We discuss the values of $U_{e3}$
and also estimate the masses of the relevant fields to realize 
an appropriate abundance of the CDM.

\newpage
\setcounter{footnote}{0}
\def\thefootnote{\arabic{footnote}}
\section{Introduction}
Recent experimental and observational results on neutrino masses \cite{nmass} 
and cold dark matter (CDM) \cite{wmap} suggest that the standard model 
(SM) should be extended by introducing some neutral fields.
A well studied candidate for the extension is the minimal 
supersymmetric SM (MSSM).
Although the MSSM contains a good CDM candidate as the lightest 
superparticle (LSP) as long as the $R$-parity is conserved,
the parameter regions preferable for the explanation of the WMAP data 
are found to be strictly restricted in certain types of the MSSM \cite{mssm}.
Confronting these situations, it seems to be interesting to consider 
models in which we can explain these new features from the same origin
in a non-supersymmetric extension of the SM: 
A certain symmetry related to the smallness of
neutrino masses can guarantee the stability of a CDM candidate.
Backgrounds that forth coming collider experiments like LHC may find
signatures of such extended models make this kind of trials 
worthy enough at present stage.
Several recent works have been done along this line \cite{nmdark}. 

In this paper we follow this line to propose an extension of 
a previously considered model by introducing a local 
U(1)$^\prime$ symmetry at TeV regions.  
As in the radiative mass generation models \cite{radmass}, we introduce an 
additional SU(2) doublet $\eta^T\equiv (\eta^+, \eta^0)$ 
to the ordinary Higgs doublet $H^T\equiv (H^+, H^0)$.
We also introduce a singlet $\phi$ whose vacuum expectation value breaks 
U(1)$^\prime$ symmetry spontaneously
down to $Z_2$ which is responsible for the stability
of the CDM candidate.
This extension seems to 
remedy defects in the previous models that certain fine tunings are 
required for both the generation of small neutrino masses and 
the reconciliation between the 
CDM abundance and the constraints from lepton flavor violating processes.
Based on such a model we calculate the value of the element
$U_{e3}$ of the MNS matrix and masses of the relevant fields 
which produce an appropriate abundance of the CDM.

\section{A model}
We consider a model with a similar symmetry to the model in 
\cite{nmdark,z2}. 
We extend it by introducing a singlet Higgs scalar $\phi$.
This extension makes the model able to contain 
an additional U(1)$^\prime$ symmetry. 
In this paper we assume that this symmetry is leptophobic and then
leptons do not have its charge for simplicity.\footnote{We need to introduce
some additional fermions to cancel gauge anomaly.  Since such extensions
will be done without changing the following results, 
we do not go further into this problem here.}
The U(1)$^\prime$ charge for the ingredients of the model is shown 
in Table 1, in which fermions are assumed to be left-handed.
Note that we need only two right-handed neutrinos 
$N_1$ and $N_2$ to generate appropriate neutrino masses and mixings 
in a minimal case.
Then the invariant Lagrangian relevant to the neutrino masses 
can be expressed as 
\begin{equation}
{\cal L}_m=\sum_{\alpha=e,\mu,\tau}
\left(h_{\alpha 1}L_\alpha H\bar N_1+h_{\alpha 2}L_\alpha\eta \bar N_2\right)
+{1\over 2}M_\ast\bar N_1^2 +{1\over 2}\lambda\phi\bar N_2^2 + {\rm h.c.}, 
\label{masslag}
\end{equation}
where we assume that Yukawa couplings for charged leptons are diagonal.
The most general invariant scalar potential up to dimension
five may also be written as
\begin{eqnarray}
V&=&{1\over 2}\lambda_1(H^\dagger H)^2 +{1\over 2}
\lambda_2(\eta^\dagger\eta)^2
+{1\over 2}\lambda_3(\phi^\dagger\phi)^2 \nonumber\\
&+&\lambda_4(H^\dagger H)(\eta^\dagger\eta)
+\lambda_5(H^\dagger\eta)(\eta^\dagger H)
+{\lambda_6\over 2M_\ast}\left[\phi(\eta^\dagger H)^2 +{\rm h.c.} 
\right]  \nonumber \\
&+& (m_H^2+\lambda_7\phi^\dagger\phi) H^\dagger H 
+(m_\eta^2+\lambda_8\phi^\dagger\phi)\eta^\dagger\eta 
+m_\phi^2\phi^\dagger\phi.
\label{pot}
\end{eqnarray}
We add a nonrenormalizable $\lambda_6$ term and a bare mass term
for $N_1$.
The scalar potential (\ref{pot}) without the $\lambda_6$ term has
an accidental U(1) symmetry, which forbids the one-loop
contribution of the $\eta$ exchange diagram to neutrino masses.
This symmetry is explicitly broken
by the Yukawa interactions (\ref{masslag}), so that
 terms like the $\lambda_6$ term, i.e. 
$(\phi^\dag \phi)^n \phi (\eta^\dag H)^2$,
 can be generated in high orders in perturbation theory in general.
 All of them contribute to radiative neutrino masses.
\footnote{It turns out that one-loop corrections 
 generating   the $\lambda_6$ term, 
 i.e. $(\phi^\dag \phi) \phi (\eta^\dag H)^2$,
 vanish if the condition (\ref{cyukawa}) discussed later is satisfied.}
Here we do not ask the origin of the $\lambda_6$ term.
They might be supposed to be effective terms generated through some
dynamics at an intermediate scale $M_\ast$.
We can check that there are no other dimension five operators
invariant under the above mentioned symmetry in the scalar potential.

\begin{figure}[t]
\begin{center}
\begin{tabular}{c|cccccccccc}
& $Q_\alpha$ & $\bar U_\alpha$ & $\bar D_\alpha$ & $L_\alpha$ 
& $\bar E_\alpha$ & $\bar N_1$ & $\bar N_2$ & $H$ & $\eta$ & $\phi$ \\\hline
U(1)$^\prime$ & $2q$ & $-2q$ & $-2q$ & 0 & 0 & 0 & $q$ & 0 &  $-q$ 
& $-2q$ \\   \hline 
$Z_2$ & +1 & +1 & +1 & +1 & +1 & +1 & $-1$ & +1 & $-1$ & +1 \\
\end{tabular}
\vspace*{3mm}

{\footnotesize {\bf Table 1.}~~ Field contents and their charges.
$Z_2$ is the residual symmetry of U(1)$^\prime$.}
\end{center}
\end{figure}

As the model discussed in \cite{nmdark}, $H$ plays the role of the ordinary
doublet Higgs scalar in the SM but $\eta$ is assumed to obtain no vacuum 
expectation value (VEV).
 A singlet scalar $\phi$ is assumed to obtain a 
VEV, which breaks U(1)$^\prime$ down to $Z_2$ ( See Table 1).
This VEV also gives the mass for $N_2$ through
$M_{N_2}=\lambda\langle\phi\rangle$ and also yields an effective coupling
for the $\lambda_6$ term as $\lambda_6\langle\phi\rangle/M_\ast$.
It can be small enough as long as $\langle\phi\rangle\ll M_\ast$ is satisfied.
Thus, the masses of the real and imaginary parts of $\eta^0$ 
are found to be almost degenerate. They are expressed as 
$M_{\eta^0}^2\simeq m_\eta^2
+(\lambda_4+\lambda_5)\langle H^0\rangle^2+\lambda_8\langle\phi\rangle^2$.
In the model discussed in \cite{nmdark}, the coupling constant
of the term corresponding to this $\lambda_6$ term 
is required to be extremely small to 
generate appropriate neutrino masses. This point is automatically 
improved by introducing the new U(1)$^\prime$ symmetry. 

\section{Masses and mixings of neutrinos}
We find that there are two
origins for the neutrino masses under these settings for the model. 
One is the ordinary seesaw mass induced
by a right-handed neutrino $N_1$ \cite{sterile} and another is 
one-loop radiative mass mediated by the exchange of $\eta^0$ and $N_2$
\cite{radmass,z2}.
These effects generate a mass matrix for three light neutrinos. 
It is expressed by
\begin{equation}
M_\nu={v^2\over M_\ast}\left[\mu^{(1)}+{\lambda_6\over 8\pi^2\lambda}
I\left({M_{N_2}^2\over M_{\eta^0}^2}\right)\mu^{(2)}\right], \qquad
I(x)={x\over 1-x}\left(1+{x\ln x\over 1-x}\right),
\label{nmass}
\end{equation}
where $v=\langle H^0\rangle$ and $\mu^{(a)}$ is defined by
\begin{equation}
\mu^{(a)}=\left(\begin{array}{ccc}
h_{ea}^2 & h_{ea}h_{\mu a} & h_{ea}h_{\tau a} \\
h_{ea}h_{\mu a} & h_{\mu a}^2 & h_{\mu a}h_{\tau a} \\
h_{ea}h_{\tau a} & h_{\mu a}h_{\tau a} & h_{\tau a}^2 \\
\end{array}\right) \quad (a=1,2).
\label{yukawa}
\end{equation}
Although two terms of $M_\nu$ may be characterized by different mass scales, 
the texture of both terms is the same as found in eq.~(\ref{yukawa}).
This type of the texture for neutrino mass matrix has been studied in 
\cite{sterile,sterile1}. 
We neglect CP phases in the following discussion.

Now we study eigenvalues and the mixing matrix for the neutrino
mass matrix (\ref{nmass}).
We consider to diagonalize $M_\nu$
by using an orthogonal matrix $U$ in such a way 
as $ U^T M_\nu U= {\rm diag}(m_1, m_2, m_3)$.
If Yukawa couplings satisfy a condition
\begin{equation}
h_{e1}h_{e2}+h_{\mu 1}h_{\mu 2}+h_{\tau 1}h_{\tau 2}
\propto [\mu^{(1)},~\mu^{(2)}]
=0,
\label{cyukawa}
\end{equation}
$\mu^{(1)}$ and $\mu^{(2)}$ can be simultaneously diagonalized. 
Since $U$ can be analytically found in such cases, 
we confine ourselves to these interesting ones.
We define a matrix $\tilde U$ as 
\begin{equation}
\tilde{U}=
\left(\begin{array}{ccc}
1 & 0 & 0 \\
0 & \cos\theta_{2} &\sin\theta_{2} \\
0 &-\sin\theta_{2} &\cos\theta_{2} \\
\end{array}\right)
\left(\begin{array}{ccc}
\cos\theta_{3} & 0 &\sin\theta_{3} \\
0 & 1 & 0 \\
-\sin\theta_{1} & 0 & \cos\theta_{3}\\
\end{array}\right).
\end{equation}
The first term of $M_\nu$ can be diagonalized by this $\tilde U$ 
if the following conditions is satisfied:
\begin{equation}
\tan\theta_2={h_{\mu 1}\over h_{\tau 1}}, \qquad 
\tan\theta_{3}={h_{e1}\over \sqrt{h_{\mu 1}^2+h_{\tau 1}^2}}.
\label{angle}
\end{equation}
Then the mass eigenvalues for the first term of $M_\nu$ are obtained by
using the following eigenvalues of $\mu^{(1)}$:
\begin{equation}
\mu_{\rm diag}^{(1)}={\rm diag}(0,~0,~h_{e1}^2
+h_{\mu 1}^2+h_{\tau 1}^2).
\label{first}
\end{equation}

We consider diagonalization of $\mu^{(2)}$ next.
At first, it should be noted that $\mu^{(2)}$ is transformed 
by the same $\tilde U$. However, if the condition (\ref{cyukawa}),
which can be written as 
\begin{equation}
h_{e2}\sin\theta_3+(h_{\mu 2}\sin\theta_2
+h_{\tau 2}\cos\theta_2)\cos\theta_3=0
\label{cond}
\end{equation}
is satisfied,
$\mu^{(2)}$ can be diagonalized by 
applying an orthogonal transformation $\tilde UU_3$ 
supplemented by an additional one given by
\begin{equation}
U_3=\left(\begin{array}{ccc}
\cos\theta_1 &\sin\theta_1 & 0  \\
-\sin\theta_1 &\cos\theta_1 & 0 \\
0 & 0 & 1 \end{array}\right).
\end{equation}
This additional transformation by $U_3$ does not affect 
the diagonalization of $\mu^{(1)}$. Consequently, 
both terms of $M_{\nu}$ can be simultaneously diagonalized 
by setting
\begin{equation}
\tan\theta_1=-{\tan\tilde\theta_2\tan\theta_2+1\over
(\tan\tilde\theta_2-\tan\theta_2)\sin\theta_3}, 
\label{ue3}
\end{equation}
where we define $\tilde\theta_2$ as $\tan\tilde\theta_2=h_{\mu 2}/h_{\tau 2}$.
Finally, we obtain nonzero mass eigenvalues of the light neutrinos as
\begin{equation}
m_2=AB~{\tan^2\theta_1+1\over \tan^2\theta_2+1}
(\tan\tilde\theta_2-\tan\theta_2)^2, \qquad
m_3={A\over 2}(\tan^2\theta_2+1)(\tan^2\theta_3+1), 
\end{equation}
where $A=2h^2_{\tau 1}v^2/M_\ast$ and 
$B=(\lambda_6/16\pi^2\lambda)(h_{\tau 2}/
h_{\tau 1})^2I(M_{N_2}^2/M_{\eta^0}^2)$.

Here we fix $\tan\theta_2=1$ which is supported by the data of 
the atmospheric neutrino and K2K experiment.  
CHOOZ experiments give the constraint on $\theta_3$ 
such as $|\sin\theta_3|<0.22$ \cite{chooz}. 
If we use these conditions, the mixing matrix $U=\tilde{U}U_3$ 
can be approximately written as 
\begin{equation}
U=\left(\begin{array}{ccc}
\cos\theta_{1} &\sin\theta_{1} & {\sin\theta_3\over\sqrt{2}}  \\
-{\sin\theta_{1}\over\sqrt{2}} &{\cos\theta_{1}\over\sqrt{2}} & 
{1\over\sqrt{2}} \\
{\sin\theta_{1}\over\sqrt{2}} &-{\cos\theta_{1}\over\sqrt{2}} & 
{1\over\sqrt{2}} \\
\end{array}\right).
\end{equation}

\input epsf.tex 
\begin{figure}[tb]
\begin{center}
\epsfxsize=7.5cm
\leavevmode
\epsfbox{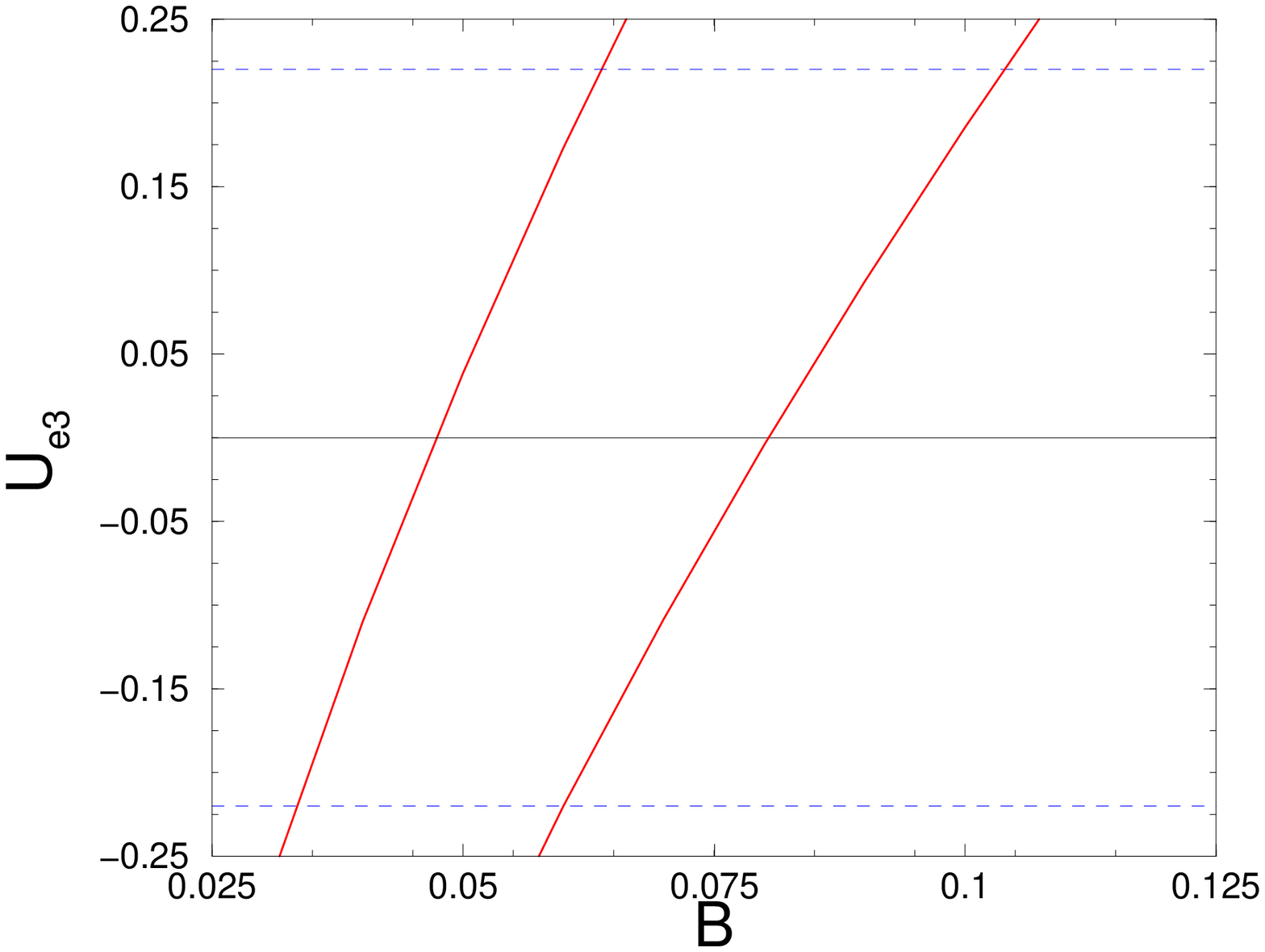}
\end{center}
{\footnotesize {\bf Fig.~1}~~$U_{e3}$ as a function of $B$. 
Allowed regions are shown as the
 regions surrounded by red solid lines.
 Horizontal dashed lines stand for the present experimental 
upper bounds for $|U_{e3}|$.}
\end{figure}

Only two mass eigenvalues $M_\nu$ are nonzero and then 
we impose that squared mass differences required by 
the neutrino oscillation data satisfy
$m_3=\sqrt{\Delta m_{\rm atm}^2}$ and 
$m_2=\sqrt{\Delta m_{\rm sol}^2}$.
Although there is a possibility that two nonzero eigenvalues have
almost degenerate values such as $\sqrt{\Delta m_{\rm atm}^2}$ and their
squared difference is given by $\Delta m^2_{\rm sol}$, we do not consider it 
since $\mu^{(1)}$ and $\mu^{(2)}$ have independent origins. 
We suppose $\theta_1=\theta_{\rm sol}$, where $\theta_{\rm sol}$ is a
mixing angle relevant to the solar neutrino.
Then we can determine $\theta_3$ through eq.~(\ref{ue3})
by using $\tan\theta_{\rm sol}$, $\sqrt{\Delta m_{\rm atm}^2}$,
$\sqrt{\Delta m_{\rm sol}^2}$ and $B$. If we use neutrino
oscillation data for these, we can find allowed regions of $\theta_3$ as 
a function of $B$.
This is shown in Fig.1, where we have used values of the measured neutrino
oscillation parameters \cite{pdg}
\begin{eqnarray}
&&\Delta m_{\rm sol}^2=8.0{^{+0.6}_{-0.4}}\times 10^{-5}~{\rm eV}^2,\quad
\Delta m_{\rm atm}^2=(1.9-3.6)\times 10^{-3}~{\rm eV}^2,\nonumber \\
&&\tan^2\theta_{\rm sol}=0.45{^{+0.09}_{-0.07}}.
\end{eqnarray}
This figure shows that $B$ is restricted in narrow regions 
such as $0.03<B<0.1$.

As an example, let us assume $M_{\eta^0}/M_{N_2}=0.3-0.7$ and then
$I(M_{N_2}^2/M_{\eta^0}^2)=0.1-1.3$.
In such cases $h_{\tau 2}/h_{\tau 1}\simeq 10(\lambda/\lambda_6)^{1/2}$ 
should be satisfied. If we obtain more constraints on the relevant coupling 
constants, we may restrict the value of $U_{e3}$ much more.
Although $U_{e3}$ takes a nonzero value for 
$0.03~{^<_\sim}~B~{^<_\sim}~0.05$ and $0.08~{^<_\sim}~B~{^<_\sim}~0.1$,
$U_{e3}=0$ is also allowed for $0.03<B<0.08$. 
The condition for the coupling constants 
can be easily satisfied even if we assume that coupling constants are $O(1)$. 
Therefore, the model needs no fine tuning to be consistent with all the 
present experimental data for neutrino oscillations.
The effective mass $m_{ee}$ for the neutrinoless double beta decay takes the values
in the range
$|m_{ee}|
~{^<_\sim}~6.3\times 10^{-3}~{\rm eV}$.

\section{Relic abundance of a CDM candidate}
The lightest field with an odd $Z_2$ charge can be stable since an even
charge is assigned to each SM content. If both the mass
and the annihilation cross section of such a field have appropriate values, 
it can be a good CDM candidate as long as it is neutral.
As found from Table~1, such candidates are $N_2$ and $\eta^0$. 
Since they have a new U(1)$^\prime$ gauge interaction,
their annihilation to quarks is considered to 
be dominantly mediated by this interaction.\footnote{A role of
U(1)$^\prime$ in annihilation of the CDM in supersymmetric models
has been studied in \cite{susy}.}
If their annihilation is
mediated only by the exchange of $\eta^0$ or $N_2$ 
through Yukawa couplings as in the model discussed in \cite{nmdark},
we cannot simultaneously explain, without fine tuning of coupling constants,
both the observed value of the CDM abundance  
and the constraints coming from lepton flavor violating processes such as 
$\mu\rightarrow e \gamma$.
Since U(1)$^\prime$ is supposed to be a generation independent gauge
symmetry, we can easily escape this problem by assuming
that the Yukawa couplings $h_{\alpha 2}$ are small enough or both $\eta^0$ and
$N_2$ are heavy enough. 
In the following study we consider the case 
that $N_2$ is lighter than $\eta^0$.
As seen in the last part of the previous section, 
this case is consistent with the
present experimental bounds for $U_{e3}$ without fine tuning.

\begin{figure}[tb]
\begin{center}
\epsfxsize=7.5cm
\leavevmode
\epsfbox{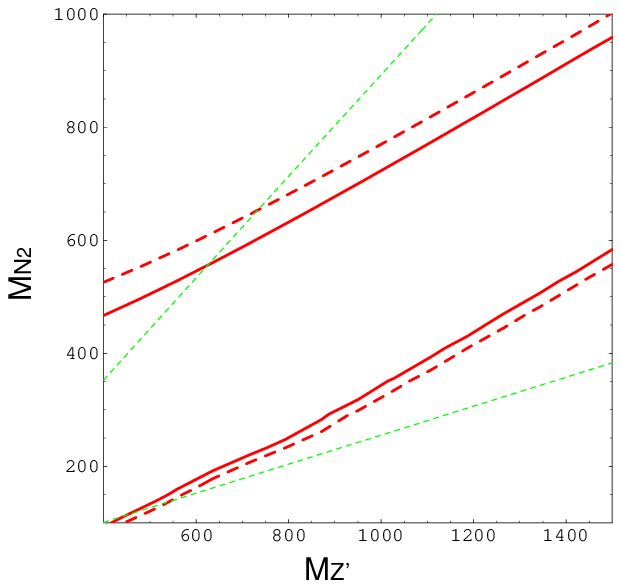}
\hspace*{5mm}
\epsfxsize=7.5cm
\leavevmode
\epsfbox{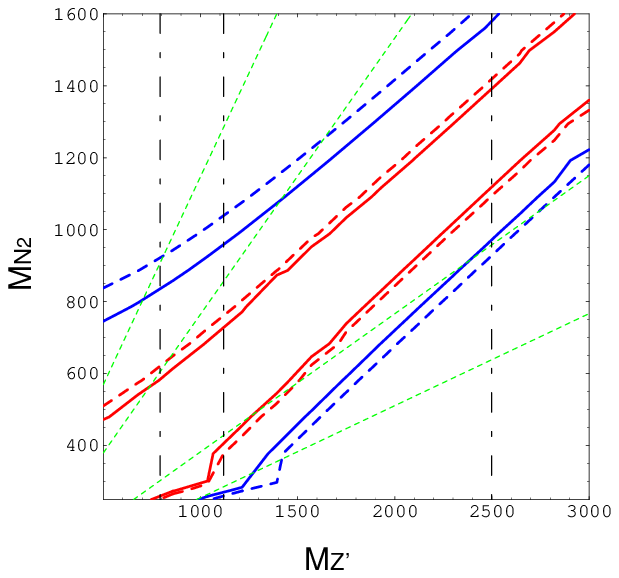}
\end{center}
{\footnotesize {\bf Fig.~2}~~Allowed regions by the WMAP data 
in the $(M_{Z^\prime}, M_{N_2})$ plane. In both figures
solid and dotted lines represent contours for $\Omega_{N_2}h^2=0.0945$ and 
$\Omega_{N_2}h^2=0.1287$. Green dotted lines stand for
 $M_{N_2}$ for typical values of $\lambda$ which are given in the text.
In the right figure vertical dash-dotted lines represent lower bounds of
 $M_{Z^\prime}$ in case of $|\theta|=10^{-2}, 5\times 10^{-3}, 10^{-3}$. }
\end{figure}
 
Now we estimate the relic abundance of $N_2$ and compare it with the
CDM abundance obtained from the WMAP data. 
We suppose that possible annihilation processes 
$N_2N_2 \rightarrow f\bar f$ 
are dominantly mediated by the U(1)$^\prime$ gauge field.
If it is expanded by relative velocity $v$ between annihilating $N_2$'s 
as $\sigma v= a + b v^2$, 
the coefficients $a$ and $b$ are expressed as
\begin{equation}
a=\sum_fc_f{g^{\prime 4}\over 2\pi}Q_{f_A}^2 q^2
{m_f^2\beta\over (s-M_{Z^\prime}^2)^2}, \qquad
b=\sum_fc_f{g^{\prime 4}\over 6\pi}(Q_{f_V}^2+Q_{f_A}^2)
q^2{M_{N_2}^2\beta \over (s-M_{Z^\prime}^2)^2},
\end{equation}
where $\beta=\sqrt{1-m_f^2/M_{N_2}^2}$ and $c_f$=3 for quarks. 
$s$ is the center of mass energy of collisions and $q$ is the U(1)$^\prime$ 
charge of $N_2$ given in Table 1. 
The charge of the final state fermion $f$ is defined as
\begin{equation}
Q_{f_V}=Q_{f_R}+Q_{f_L}, \qquad Q_{f_A}=Q_{f_R}-Q_{f_L}.
\end{equation}
Using these quantities, the present relic abundance of $N_2$ can be 
estimated as \cite{cdm},
\begin{equation}
\Omega_{N_2} h^2|_0=
\left.{M_{N_2} 
n_{N_2}\over \rho_{\rm cr}/h^2 }\right|_0
\simeq{8.76\times 10^{-11}g_\ast^{-1/2}x_F\over 
(a+3b/x_F)~{\rm GeV}^2 }.
\end{equation} 
where $g_\ast$ enumerates the degrees of freedom of relativistic 
fields at the freeze-out temperature $T_F$ of $N_2$.
$T_F$ is determined through the equation for a dimensionless parameter
 $x_F=M_{N_2}/T_F$
\begin{equation}
x_F=\ln{0.0955m_{\rm pl}M_{N_2}(a+6b/x_F)
\over (g_\ast x_F)^{1/2}},
\end{equation}  
where $m_{\rm pl}$ is the Planck mass. 
If we fix the U(1)$^\prime$ charge of fields and  
its coupling constant $g^\prime$, we can estimate 
the present $N_2$ abundance using these
formulas. Assuming a GUT relation $g^\prime=\sqrt{5/3}g_1$ 
and $q=0.6$ as an example, we calculate $\Omega_{N_2}h^2$.
The results are given in Fig.~2.

In the left figure of Fig.~2 we plot favorable regions in 
the $(M_{Z^\prime}, M_{N_2})$ plane, 
where $\Omega_{N_2}h^2$ takes values in the range 0.0945 -- 0.1285, which is
required by the WMAP data. 
$\Omega_{N_2}h^2$ has a valley in the parameter region of Fig.~2, 
and therefore the allowed regions appear as two narrow bands,
each sandwiched by a solid line and a dashed line. 
Since $M_{N_2}$ and $M_{Z^\prime}$ are induced by $\langle\phi\rangle$
and written as
\begin{equation}
M_{N_2}=\lambda\langle\phi\rangle, \qquad
M_{Z^\prime}=2\sqrt{2}g^\prime q\langle\phi\rangle,
\end{equation}
$M_{N_2}$ is determined by $M_{Z^\prime}$. 
We plot this $M_{N_2}$ values by green dotted lines for $\lambda=0.2$ and 0.7.
The lower bounds of $M_{Z^\prime}$ come from constraints for $ZZ^\prime$
mixing and direct search of $Z^\prime$.
$H$ is assumed to have no U(1)$^\prime$ charge and then 
its VEV induces no $ZZ^\prime$ mixing. Moreover, since it is leptophobic,
the constraints on $M_{Z^\prime}$ obtained from its hadronic decay 
is rather weak. Thus, the lower bounds of $M_{Z^\prime}$ may be 
$M_{Z^\prime}~{^>_\sim}~450$~GeV in the present model \cite{leptph}. 
Taking account of this, Fig.~1 shows that this model can well explain 
the CDM abundance.
Since $\lambda$ is included in the definition of $B$, 
values of $\theta_3$ may be constrained by the mass of the CDM 
if we can obtain more informations on
$\lambda_6$, $h_{\tau 2}/h_{\tau 1}$ and $M_{Z^\prime}$.

Here we briefly discuss the relation to lepton flavor violating 
processes such as $\mu\rightarrow e\gamma$. 
As in the model of \cite{mueg}, $\mu\rightarrow e\gamma$
is induced through the mediation of $\eta^0$ and $N_2$. 
Its branching ratio can be given by 
\begin{eqnarray}
&&B(\mu\rightarrow e\gamma)={3\alpha\over 64\pi(G_FM_{\eta^0}^2)^2}
\left|h_{\mu 2}h_{e2}F_2\left({M_{N_2}^2\over M_{\eta^0}^2}\right)\right|^2,
\nonumber  \\
&&F_2(x)={1\over 6(1-x)^4}(1-6x+3x^2+2x^3-6x^2\ln x).
\end{eqnarray}  
Taking account that $1/12<F_2(x)<1/6$ is satisfied in case of 
$M_{N_2}<M_{\eta^0}$ and imposing the present experimental upper bound
$B(\mu\rightarrow e\gamma)~{^<_\sim}~1.2\times 10^{-11}$, we find that
$M_{\eta^0}$ should satisfy
\begin{equation}
M_{\eta^0}~{^>_\sim}~(360-500)\left({h_{\tau 2}\over 0.1}\right)
~{\rm GeV}.
\end{equation}
Here we use the results of the previous section.
Constraints coming from $\mu\rightarrow e\gamma$ and the 
CDM abundance can be consistent for reasonable values of  $h_{\tau 2}$.
Since $N_2$ annihilation due to an $\eta^0$ exchange is
ineffective for these values of couplings and masses \cite{nmdark}, 
the results of the $N_2$ abundance given above is not affected by 
this process. 
 
Finally, it may be useful to refer to the cases of general U(1)$^\prime$.
In these cases a crucial condition for the mass 
of the U(1)$^\prime$ gauge field comes from the constraint 
for $ZZ^\prime$ mixing.
A mass matrix for neutral gauge bosons can be expressed as
\begin{equation}
\left(\begin{array}{cc}
{1\over 2}(g_1^2+g_2^2)v^2 & -g^\prime \sqrt{g_1^2+g_2^2}q_Hv^2 \\
-g^\prime \sqrt{g_1^2+g_2^2}q_H v^2 & 
2g^{\prime 2} q_\phi^2\left( 4\langle\phi\rangle^2+ v^2\right) 
\end{array}\right),
\label{z2mass}
\end{equation}
where $q_H$ and $q_\phi$ stand for the U(1)$^\prime$ charge of $H$ and 
$\phi$.   
Since a $ZZ^\prime$ mixing angle $\theta$ is known 
to be strongly suppressed \cite{mixing},
the magnitude of $\langle\phi\rangle$ should satisfy
\begin{equation}
\langle\phi\rangle ~{^>_\sim}~ {v\over 2}
{(g_1^2+g_2^2)^{1/4} \over (2qg^\prime\vert\theta\vert)^{1/2}}.
\label{gmix}
\end{equation}
This condition gives a lower bound on both $M_{Z^\prime}$
and $M_{N_2}$.
In the right panel of Fig.~2 we plot this bound in cases of 
$|\theta|=10^{-2}, 5\times 10^{-3}, 10^{-3}$,
which are drawn by vertical dash-dotted lines.
We also plot the values of $M_{N_2}$ for $\lambda=0.2, 0.3, 0.6$ and
0.9. They are drawn by green dotted lines.
Although $|\theta|$ should be less than $10^{-3}$, we may suppose larger
values of $|\theta|$ in eq.~(\ref{gmix}) by extending 
the model without changing the results in the previous section.
In fact, if the model has two Higgs doublets $H_u$ and $H_d$ 
which couple to up- and down-sectors respectively, off-diagonal 
elements of eq.~(\ref{z2mass}) is proportional to 
$g^\prime(q_{H_u}\langle H_u\rangle^2 -q_{H_d}\langle H_d\rangle^2)$ 
where $q_{H_{u,d}}$ expresses the U(1)$^\prime$ charge.  
Cancellation between these two contributions can make
the $ZZ^\prime$ mixing smaller for the same value of 
$\langle\phi\rangle$. In such cases we can apply 
this effect by using larger $|\theta|$ values in eq.~(\ref{gmix}).
In this figure $\theta$ values smaller than $10^{-3}$ should be 
understood based on this reasoning.

On the other hand, the introduction of additional Higgs doublets may require
us to take account of new final states for the $N_2$ annihilation 
induced by the $Z^\prime$ exchange. 
If $N_2$ is heavier than $W^\pm$, the final states should include gauge bosons 
and Higgs scalars such as $W^+W^-$, $H_i^0H_j^0$, $W^\pm H^\mp$,
$H^+H^-$ and $ZH^0_i$, where $H_i^0$ is a mass eigenstate of the neutral Higgs.
Since the annihilation to $W^+W^-$ is suppressed by the $ZZ^\prime$ 
mixing in the present model, important modes are expected to be $H_i^0H_j^0$ 
and they may give the same order of contributions as the annihilation to 
$f\bar f$ \cite{cdm}.
In order to take such effects into account
without practicing tedious estimation of such processes, 
we show in the right figure of Fig.~2 an additional $\Omega_{N_2}h^2$ 
contour which is obtained by using $5\times(\sigma v)_{f\bar f}$
for cross section. It is drawn by blue lines.
 An original contour for the cross section 
$(\sigma v)_{f\bar f}$ is drawn by red lines.\footnote{
In this calculation we use the U(1)$^\prime$ charge
assignment for Higgs doublets, quarks and leptons such as 
$q_{H_u}=q_{H_d}=-2q,~q_Q=q_L=2q$. }
Since main parts of the cross section into these final states are
expected to have the similar dependence on $M_{Z^\prime}$ and $M_{N_2}$, 
this is considered to give good references for these cases. 
This figure also suggests that this kind of models can explain the CDM
abundance even under the constraint for $Z^\prime$ physics. 

\section{Summary}
We have studied neutrino masses and CDM abundance 
in a non-supersymmetric, but  U(1)$^\prime$ symmetric model which is 
obtained from the SM by adding certain neutral fields. 
Neutrino masses are generated through both the seesaw mechanism 
with a single right-handed neutrino and the one-loop radiative 
effects. They induce the same texture which can realize favorable mass 
eigenvalues and mixing angles. One of the introduced neutral fields is
stable due to an unbroken $Z_2$ symmetry
 which is the residual symmetry of  the spontaneously broken
U(1)$^\prime$.
Thus it can be a good CDM candidate. Since it has the
U(1)$^\prime$ gauge interaction, the annihilation is dominantly 
mediated through this
interaction. If this U(1)$^\prime$ symmetry is broken at a suitable 
scale, the present relic abundance of right-handed neutrinos 
can explain the WMAP result for the CDM abundance.
This model suggests that two of the biggest questions in the SM, that
is, neutrino masses and the CDM 
may be explained on the common basis of an extension of the SM. 
An interesting feature of the model is that the
value of the third mixing angle $\theta_3$ may be related to 
the mass of the CDM.
The model may be examined through the search of the $Z^\prime$ and
the additional Higgs doublet $\eta$ at LHC.

\section*{Acknowledgement}
This work is partially supported by a Grant-in-Aid for Scientific
Research (C) from Japan Society for Promotion of Science (No.17540246
and No.18540257).
We would like to thank the referee of Physics Letters B
for pointing out that the originally proposed $Z_2$ is redundant
because  $Z_2$ is nothing but the residual symmetry of U(1)$^\prime$.

\end{document}